\documentstyle[prd,aps]{revtex}
\begin{document}
\draft
%*
%\twocolumn[\hsize\textwidth\columnwidth\hsize\csname @twocolumnfalse\endcsname

\title
{Fractional spin and the Pauli term}

\author {C. R. Hagen\cite{Hagen}}

\address
{Department of Physics and Astronomy\\
University of Rochester\\
Rochester, N.Y. 14627}

\maketitle

\begin{abstract}
It has recently been claimed that the inclusion of a Pauli term in (2+1)
dimensions gives rise to a new type of anomalous spin term.  The form of that
term is shown to contradict the structure relations for the inhomogeneous
Lorentz group. 
\end{abstract}

%*
\vskip2pc

In 1984 [1] it was shown that it is possible to formulate a pure Chern-Simons
gauge theory.  That theory was found to have the peculiar feature that the
commutator of the angular momentum operator $L$ with the fundamental charge 
field has an anomalous term (i.e., a fractional spin) proportional to to the 
product of that field with the charge operator $Q$.  More recently it has been 
claimed [2] that if a Pauli-type coupling is included, an additional anomalous 
spin term is induced.  Specifically, it is asserted that the commutator of $L$
with the scalar charge field $\phi$ has the form
$$[L,\phi(y)]=-i({\bf y}\times{\bf \nabla})\phi(y)-{e^2\over 2\pi \kappa}Q\phi
(y)+i{g\over 2}{\bf y}\cdot{\bf E}\phi(y)$$
where $e$, $g$, and $\kappa$ are constants.  It is shown here that such a term
is inconsistent with the underlying Poincar\'e algebra in (2+1) dimensions.

The proof of this result follows immediately upon inserting the above equation
between the operators $e^{-i{\bf P}\cdot {\bf a}}$ and $e^{i{\bf P}\cdot {\bf
a}}$ where  ${\bf a}$ is an arbitrary two-vector.  Using the result
$$A({\bf x} + {\bf a}) =  e^{-i{\bf P}\cdot {\bf a}}A({\bf x})e^{i{\bf P}\cdot {\bf a}}$$
for an arbitrary operator $A$ together with the structure relation
$$[L,P^i]=i\epsilon^{ij}P_j,$$
this is seen to yield
$$[L,\phi( y)]=-i({\bf y}\times {\bf \nabla})\phi(y)-{e^2\over 2\pi
\kappa}Q\phi(y)+i{g\over 2}({\bf y}-{\bf a})\cdot {\bf E}\phi(y)$$
upon letting ${\bf y} \rightarrow
 {\bf y}-{\bf a}$.   This leads to the immediate result
that the relation claimed in ref. 2 cannot be valid for nonzero $g$.

\acknowledgments

This work is supported in part by the U.S. Department of Energy Grant
No.DE-FG02-91ER40685.

\end{document}